\documentclass[a4paper,twoside]{article}

\usepackage{ifthen}
\newboolean{adp}
\setboolean{adp}{false}
\newboolean{arxiv}
\setboolean{arxiv}{true}
\newboolean{notadp}
\setboolean{notadp}{true}
\ifadp
\setboolean{notadp}{false}
\fi
\newboolean{notarxiv}
\setboolean{notarxiv}{true}
\ifarxiv
\setboolean{notarxiv}{false}
\fi

\usepackage{times}

\usepackage[american]{babel}

\ifnotadp
\usepackage{a4wide}
\input{fleqn.clo}
\PassOptionsToPackage{fleqn}{amsmath}
\PassOptionsToPackage{fleqn}{amstex}
\fi

\usepackage{amsmath}
\usepackage{amsfonts}
\usepackage{amssymb}
\allowdisplaybreaks

\ifnotarxiv
\usepackage{xcolor}
\xdefinecolor{mylinkcolor}{rgb}{0,0,0.5}
\usepackage[
	bookmarksnumbered, bookmarksopen, bookmarksopenlevel=2,
	breaklinks=true, colorlinks=true,
	filecolor=mylinkcolor, citecolor=mylinkcolor, linkcolor=mylinkcolor,
	urlcolor=mylinkcolor, menucolor=mylinkcolor,
]{hyperref}
\fi

\ifarxiv 
\usepackage{xcolor}
\xdefinecolor{mylinkcolor}{rgb}{0,0,0.5}
\usepackage[
	bookmarksnumbered, bookmarksopen, bookmarksopenlevel=1,
	colorlinks=true,
	filecolor=mylinkcolor, citecolor=mylinkcolor, linkcolor=mylinkcolor,
	urlcolor=mylinkcolor, menucolor=mylinkcolor,
]{hyperref}
\fi

\ifnotadp
\usepackage[merge,square,comma,numbers,sort&compress]{natbib}
\fi
\newcommand{\mycite}[2]{
\ifnotadp\cite{#2}\fi%
\ifadp\cite{#1}\fi}

\providecommand{\inst}[1]{$\,^{#1}$}

\def\vct#1{\mathbf{#1}}

\def\nln{\nonumber\\ & \quad}
\def\nlqn{\nonumber\\ & \quad \qquad}

\def\nunit{n}

\newcommand{\vnunit}{\vct{\nunit}_{12}}

\newcommand{\spin}[3]{\hat{S}_{#1\, (#2)(#3)}}
\def\canmom{P}
\newcommand{\mom}[2]{{\canmom}_{#1\,#2}}

\newcommand{\vmom}[1]{{\vct{\canmom}}_{#1}}
\newcommand{\vnxa}[1]{{\vct{n}}_{#1}}
\newcommand{\vnun}{{\vnunit}}

\newcommand{\scpm}[2]{(#1\,#2)}
\newcommand{\vspin}[1]{\hat{\vct{S}}_{#1}}
\newcommand{\vx}[1]{\hat{\vct{z}}_{#1}}

\def\gravthree{G}
\def\dim{d}
\def\rel{r_{12}} 

\begin{document}

\def\tit{Next-to-next-to-leading order post-Newtonian spin(1)-spin(2) Hamiltonian for self-gravitating binaries}
\def\titleShort{NNLO spin(1)-spin(2) Hamiltonian for self-gravitating binaries}

\def\Fauthor{Johannes Hartung}
\def\FauthorShort{J.\ Hartung}
\def\Sauthor{Jan Steinhoff}
\def\SauthorShort{J.\ Steinhoff}

\def\FauthorFootnote{Corresponding author\quad
	E-mail:~\textsf{johannes.hartung@uni-jena.de},
	Phone: +49\,3641\,947\,106,
	Fax: +49\,3641\,947\,102}

\def\FauthorInst{1}
\def\SauthorInst{1,2}


\def\Fadr{Theoretisch--Physikalisches Institut, \\
	Friedrich--Schiller--Universit\"at, \\
	Max--Wien--Platz 1, 07743 Jena, Germany, EU}

\def\Sadr{Centro Multidisciplinar de Astrof\'isica (CENTRA), Departamento de F\'isica, \\
	Instituto Superior T\'ecnico (IST), Universidade T\'ecnica de Lisboa, \\ 
	Avenida Rovisco Pais 1, 1049-001 Lisboa, Portugal, EU}

\def\abs{We present the next-to-next-to-leading order post-Newtonian (PN)
spin(1)-spin(2) Hamiltonian for two self-gravitating spinning compact objects. If \emph{both} 
objects are rapidly rotating, then the corresponding interaction
is comparable in strength to a 4PN effect. The Hamiltonian is checked 
via the global Poincar\'e algebra with
the center-of-mass vector uniquely determined by an ansatz.}

\def\pacs{04.25.Nx, 04.20.Fy, 04.25.-g, 97.80.-d, 45.50.Jf}
\def\keyw{post-Newtonian approximation; canonical formalism;
	approximation methods; equations of motion; binary stars}

\def\ack{
We thank G. Sch\"afer for suggesting this interesting research project, for very useful
and encouraging discussions, and for strongly supporting this work.
We also thank P.\ Jaranowski for sharing his insight in the calculation of
the 3PN point-mass Hamiltonian and for providing several test integrals.
We additionally thank M. Levi for pleasant collaboration on the comparison of our result 
with the potential obtained within the EFT approach, which will be published soon.
We further gratefully acknowledge useful discussions 
with D.\ Brizuela on {\scshape xPert} and doing perturbation theory in 
arbitrary dimensions,
with M.\ Q.\ Huber on three-body integral related Appell $F_4$ functions,
with S.\ Hergt on the Poincar\'e algebra and center-of-mass vector ans\"atze,
with A.\ Nagar on the inclusion of spin into the EOB approach,
with H.\ Witek on numerical relativity, 
with M.\ Tessmer and J.\ Sperrhake on the manuscript and irreducible algebraic combinations of spin(1)-spin(2) structure. 
This work is funded by the Deutsche Forschungsgemeinschaft (DFG) through
the Research Training Group GRK 1523 ``Quanten- und Gravitationsfelder'' and
the Collaborative Research Center SFB/TR7 ``Gravitations\-wellen\-astronomie,''
and by the FCT (Portugal) through PTDC project CTEAST/098034/2008.}

\ifnotadp
\title{\tit}
\author{{\Large \Fauthor\thanks{\FauthorFootnote} \inst{\FauthorInst} and
	\Sauthor\inst{\SauthorInst} } \medskip \\
	\inst{1} \Fadr \\
	\inst{2} \Sadr}
\maketitle
\begin{abstract}
\abs
\end{abstract}
\noindent PACS numbers: \pacs 
\fi

\ifadp
\title[\titleShort]{\tit}
\author[\FauthorShort]{\Fauthor\footnote{\FauthorFootnote}\inst{\FauthorInst}}
\author[\SauthorShort]{\Sauthor\inst{\SauthorInst}}
\address[\inst{1}]{\Fadr}
\address[\inst{2}]{\Sadr}
\begin{abstract}
\abs
\end{abstract}
\maketitle
\fi

\ifnotarxiv
\hypersetup{pdftitle={\tit}, pdfauthor={\Fauthor, \Sauthor}}
\fi

%
%

\section{Introduction}\label{sec:intro}
In the present article the next-to-next-to-leading order (NNLO) post-Newtonian (PN)
spin(1)-spin(2) Hamiltonian for two self-gravitating spinning compact objects is
derived. This Hamiltonian is of the order 4PN if \emph{both} objects
are rapidly rotating. The present article is a continuation of \cite{Hartung:Steinhoff:2011:1}.

Spin(1)-spin(2) coupling in the PN approximation to general relativity was tackled by various
authors during the last decades. The leading order interaction was calculated, e.g., in
\cite{DEath:1975} with classical spins and in \cite{Barker:OConnell:1975,Barker:OConnell:1979}
with quantum mechanical spins. A canonical treatment of the next-to-leading order was done in \mycite{Steinhoff:Schafer:Hergt:2008}{Steinhoff:Schafer:Hergt:2008, *Steinhoff:Hergt:Schafer:2008:2}
(and its $n$-body extension in \cite{Hartung:Steinhoff:2010}) via the canonical formalism of
Arnowitt, Deser, and Misner \mycite{Arnowitt:Deser:Misner:1962}{Arnowitt:Deser:Misner:1962, *Arnowitt:Deser:Misner:2008}
enhanced from point-masses to linear order in spin in \cite{Steinhoff:Schafer:2009:2, Steinhoff:2011}.
This formalism was also used to derive the Hamiltonian presented in this article.
There were also several noncanonical approaches for the next-to-leading order, namely
\mycite{Porto:Rothstein:2008:1, Levi:2008}{Porto:Rothstein:2008:1, *Porto:Rothstein:2008:1:err, Levi:2008}
(and an incomplete result in \cite{Porto:Rothstein:2006})
which calculated the spin(1)-spin(2) interaction in the effective field theory formalism.
For further literature on spin interactions within the PN approximation see \cite{Hartung:Steinhoff:2011:1}.

Unfortunately the 4PN point-mass Hamiltonian is not known yet. Thus the Hamiltonian
obtained in the present article is currently not very useful within the Taylor-expanded
post-Newtonian series, even if both objects are rapidly rotating. Further,
the Hamiltonian is at most comparable in size to a 4PN effect, so it is particularly interesting to consider its
effect on the motion of compact binaries during the \emph{very} late inspiral phase.
However, during this phase the PN approximation will become increasingly inadequate
due to the highly nonlinear behavior of the dynamics. To overcome this problem it is most convenient
to extrapolate to this nonlinear regime by resumming the PN series. Such a resummation
was successfully implemented into the effective-one-body (EOB) approach, see, e.g.,
\cite{Buonanno:Damour:1999,Buonanno:Damour:2000,Damour:Iyer:Nagar:2009,Damour:Nagar:2009:2,Buonanno:etal:2009},
which analytically provides complete binary inspiral gravitational waveforms that are in good agreement with numerical relativity.
As the parameter space of \emph{spinning} binaries is very large, it is invaluable
to have such analytic methods at hand for the creation of waveform template banks
to be used in future gravitational wave astronomy. For the same reason spin-dependent
PN Hamiltonians are expected to be important for calibrating the EOB approach, whereas
for the spin-independent part a calibration to numerical relativity already works
reasonably well \cite{Damour:Nagar:2009:2, Buonanno:etal:2009} (also for the
nonprecessing spinning case \cite{Pan:etal:2009}).
In order to further improve the accuracy of the EOB approach for the spinning case,
the Hamiltonian derived in the present article should be valuable. Some of the
spin-dependent PN Hamiltonians mentioned in \cite{Hartung:Steinhoff:2011:1} were already implemented in the EOB approach
\cite{Damour:2001,Damour:Jaranowski:Schafer:2008:3,Barausse:Buonanno:2009,Nagar:2011,Barausse:Buonanno:2011},
see also \cite{Pan:etal:2009}. Notice that this even includes the NNLO spin-orbit
Hamiltonian \cite{Nagar:2011,Barausse:Buonanno:2011} obtained only very recently
in our previous article \cite{Hartung:Steinhoff:2011:1}.
But at the PN spin(1)-spin(2) level only the leading order Hamiltonian
was incorporated into the EOB approach yet, though an extension to higher order spin(1)-spin(2) couplings
is in principle possible \cite{Damour:2001,Barausse:Buonanno:2009}.
Notice that the EOB Hamiltonians in \cite{Barausse:Buonanno:2009,Barausse:Buonanno:2011} exactly implements
the test-spin Hamiltonian in a Kerr background \cite{Barausse:Racine:Buonanno:2009}
and thus the corresponding spin(1)-spin(2) coupling through all PN orders.

In a forthcoming publication we will provide much more details on the calculation of the Hamiltonian
in the present article and of the one in \cite{Hartung:Steinhoff:2011:1} as well.
A comparison of the results given in this article to the recently obtained NNLO spin(1)-spin(2) 
potential calculated within an EFT approach \cite{Levi:2011} will be postponed to 
a later publication due to the very complicated calculations necessary for the conversion.

The article is organized as follows. The next-to-next-to-leading order spin(1)-spin(2)
Hamiltonian is presented in Sect.\ \ref{sec:result}. 
The Hamiltonian is checked via
the global Poincar\'e algebra in Sect.\ \ref{sec:poincare}, where the
center-of-mass vector is uniquely determined from an ansatz. 

Three-dimensional vectors are written in boldface and their components are
denoted by Latin indices. The scalar product between two vectors $\vct{a}$ and 
$\vct{b}$ is denoted by $(\vct{a}\vct{b}) \equiv (\vct{a} \cdot \vct{b})$.
Our units are such that $c=1$. There is no special convention for Newton's gravitational
constant $\gravthree$. In the results $\vmom{a}$ denotes the canonical linear momentum of the $a$th 
object, $\hat{\vct{z}}_a$ the canonical conjugate position of the object, $m_a$ the mass of the object, 
$\hat{\vct{S}}_a$ and $\spin{a}{i}{j}$ the spin vector and the spin tensor of the object, 
$r_{ab}=|\hat{\vct{z}}_a - \hat{\vct{z}}_b|$ the relative distance between two objects, 
and $\vnxa{ab} = (\hat{\vct{z}}_a - \hat{\vct{z}}_b)/r_{ab}$ 
the direction vector pointing from object $b$ to object $a$.
In the binary case the object labels $a, b$ take only the values $1$ and $2$.
The round brackets around the indices of the canonical spin tensor
$\spin{a}{i}{j}$ indicate that its components are given in a local Lorentz
basis, which is essential for the canonical formalism, see \cite{Steinhoff:Schafer:2009:2, Steinhoff:2011}.

\section{Result}\label{sec:result}
The derivation of the result followed along the same lines as in the spin-orbit case \cite{Hartung:Steinhoff:2011:1}.
In particular we used the free {\scshape Mathematica} \cite{Wolfram:2003} package {\scshape xTensor} \cite{MartinGarcia:2002} for all computations, especially because of its fast index
canonicalizer based on the package {\scshape xPerm} \cite{MartinGarcia:2008}. We also used
the package {\scshape xPert} \cite{Brizuela:MartinGarcia:MenaMarugan:2009}, which is part of {\scshape xTensor},
for performing the perturbative part of our calculations. Furthermore we wrote several 
{\scshape Mathematica} packages ourselves for evaluating integrals.
It turns out after using the integration procedures mentioned in \cite{Hartung:Steinhoff:2011:1} that all 
integrals of the generalized Riesz-type appearing at spin(1)-spin(2) level can be reduced to Gamma functions and
Polygamma functions, which can be handled without any problems by {\scshape Mathematica}.
Further the $\dim$-dimensional UV-analysis described in \cite{Damour:Jaranowski:Schafer:2001, Damour:Jaranowski:Schafer:2008:2}
and in \cite{Hartung:Steinhoff:2011:1} gave contributions to intermediate expressions like in the
spin-orbit case, however they again exactly canceled in the final result. 

The next-to-next-to-leading order spin(1)-spin(2) Hamiltonian we obtained as a result of the procedures discussed 
in \cite{Hartung:Steinhoff:2011:1} is given by
\begin{align}
  H^{\text{NNLO}}_{\text{SS}} & = \frac{\gravthree^3}{\rel^5}\biggl[
	-\scpm{\vspin{1}}{\vspin{2}}\left(
		\frac{63}{4} m_1^2
		+\frac{145}{8} m_1 m_2
	\right) 
	+ \scpm{\vnun}{\vspin{1}}\scpm{\vnun}{\vspin{2}}\left(
		\frac{105}{4} m_1^2
		+\frac{289}{8} m_1 m_2
	\right)
\biggr]\nln
 + \frac{\gravthree^2}{\rel^4}\biggl[
	((\vnun \times \vmom{1})\,\vspin{1})((\vnun \times \vmom{1})\,\vspin{2})\biggl(
		\frac{12}{m_1}
		+\frac{9 m_2}{m_1^2}
	\biggr) \nlqn
	-\frac{81}{4 m_1}((\vnun \times \vmom{2})\,\vspin{1})((\vnun \times \vmom{1})\,\vspin{2})
	-\frac{27}{4 m_1}((\vnun \times \vmom{1})\,\vspin{1})((\vnun \times \vmom{2})\,\vspin{2})
	\nlqn
	-\frac{5}{2 m_1}\scpm{\vmom{1}}{\vspin{1}}\scpm{\vmom{2}}{\vspin{2}}
	+\frac{29}{8 m_1}\scpm{\vmom{2}}{\vspin{1}}\scpm{\vmom{1}}{\vspin{2}}
	-\frac{21}{8 m_1}\scpm{\vmom{1}}{\vspin{1}}\scpm{\vmom{1}}{\vspin{2}}
	\nlqn
	+\scpm{\vnun}{\vspin{1}}\scpm{\vmom{1}}{\vspin{2}}\biggl\{
		\left(\frac{33}{2 m_1} + \frac{9 m_2}{m_1^2}\right)\scpm{\vnun}{\vmom{1}}
		-\left(\frac{14}{m_1} + \frac{29}{2 m_2}\right)\scpm{\vnun}{\vmom{2}}
	\biggr\} \nlqn
	+\scpm{\vmom{1}}{\vspin{1}}\scpm{\vnun}{\vspin{2}}\biggl\{
		\frac{4}{m_1}\scpm{\vnun}{\vmom{1}}
		-\left(\frac{11}{m_1} + \frac{11}{m_2}\right)\scpm{\vnun}{\vmom{2}}
	\biggr\} \nlqn
	+\scpm{\vnun}{\vspin{1}}\scpm{\vnun}{\vspin{2}}\biggl\{
		-\frac{12}{m_1} \scpm{\vnun}{\vmom{1}}^2
		-\frac{10}{m_1} \vmom{1}^2
		+\frac{37}{4 m_1}
			\scpm{\vmom{1}}{\vmom{2}} \nlqn
		+\frac{255}{4 m_1}
			\scpm{\vnun}{\vmom{1}}\scpm{\vnun}{\vmom{2}}
	\biggr\} 
	+\scpm{\vspin{1}}{\vspin{2}}\biggl\{
		-\left(\frac{25}{2 m_1} + \frac{9 m_2}{m_1^2}\right) \scpm{\vnun}{\vmom{1}}^2
		+ \frac{49}{8 m_1} \vmom{1}^2 \nlqn
		+ \frac{35}{4 m_1}
			\scpm{\vnun}{\vmom{1}}\scpm{\vnun}{\vmom{2}} 
		- \frac{43}{8 m_1}
			\scpm{\vmom{1}}{\vmom{2}}
	\biggr\}
\biggr]\nln
 + \frac{\gravthree}{\rel^3}\biggl[
	\frac{((\vmom{1} \times \vmom{2})\,\vspin{1})((\vmom{1} \times \vmom{2})\,\vspin{2})}{16 m_1^2 m_2^2}
	-\frac{9 ((\vmom{1} \times \vmom{2})\,\vspin{1})((\vnun \times \vmom{2})\,\vspin{2})\scpm{\vnun}{\vmom{1}}}{8 m_1^2 m_2^2} \nlqn
	-\frac{3 ((\vnun \times \vmom{2})\,\vspin{1})((\vmom{1} \times \vmom{2})\,\vspin{2})\scpm{\vnun}{\vmom{1}}}{2 m_1^2 m_2^2} \nlqn
	+((\vnun \times \vmom{1})\,\vspin{1})((\vnun \times \vmom{1})\,\vspin{2})\biggl(
		\frac{9 \vmom{1}^2}{8 m_1^4}
		+ \frac{15 \scpm{\vnun}{\vmom{2}}^2}{4 m_1^2 m_2^2}
		- \frac{3 \vmom{2}^2}{4 m_1^2 m_2^2}
	\biggr) \nlqn
	+((\vnun \times \vmom{2})\,\vspin{1})((\vnun \times \vmom{1})\,\vspin{2})\biggl(
		-\frac{3 \vmom{1}^2}{2 m_1^3 m_2}
		+\frac{3 \scpm{\vmom{1}}{\vmom{2}}}{4 m_1^2 m_2^2} \nlqn
		-\frac{15 \scpm{\vnun}{\vmom{1}}\scpm{\vnun}{\vmom{2}}}{4 m_1^2 m_2^2}
	\biggr) 
	+((\vnun \times \vmom{1})\,\vspin{1})((\vnun \times \vmom{2})\,\vspin{2})\biggl(
		\frac{3 \vmom{1}^2}{16 m_1^3 m_2} \nlqn
		-\frac{3 \scpm{\vmom{1}}{\vmom{2}}}{16 m_1^2 m_2^2} 
		-\frac{15 \scpm{\vnun}{\vmom{1}}\scpm{\vnun}{\vmom{2}}}{16 m_1^2 m_2^2}
	\biggr) \nlqn
	+ \scpm{\vmom{1}}{\vspin{1}}\scpm{\vmom{1}}{\vspin{2}}\biggl(
		\frac{3 \scpm{\vnun}{\vmom{2}}^2}{4 m_1^2 m_2^2} 
		- \frac{\vmom{2}^2}{4 m_1^2 m_2^2}
	\biggr) \nlqn
	+ \scpm{\vmom{1}}{\vspin{1}}\scpm{\vmom{2}}{\vspin{2}}\biggl(
		-\frac{\vmom{1}^2}{4 m_1^3 m_2}
		+\frac{\scpm{\vmom{1}}{\vmom{2}}}{4 m_1^2 m_2^2}
	\biggr) \nlqn
	+ \scpm{\vmom{2}}{\vspin{1}}\scpm{\vmom{1}}{\vspin{2}}\biggl(
		\frac{5\vmom{1}^2}{16 m_1^3 m_2}
		-\frac{3\scpm{\vmom{1}}{\vmom{2}}}{16 m_1^2 m_2^2}
		-\frac{9\scpm{\vnun}{\vmom{1}}\scpm{\vnun}{\vmom{2}}}{16 m_1^2 m_2^2}
	\biggr) \nlqn
	+ \scpm{\vnun}{\vspin{1}}\scpm{\vmom{1}}{\vspin{2}}\biggl(
		\frac{9 \scpm{\vnun}{\vmom{1}} \vmom{1}^2}{8 m_1^4}
		-\frac{3 \scpm{\vnun}{\vmom{2}} \vmom{1}^2}{4 m_1^3 m_2}
		-\frac{3 \scpm{\vnun}{\vmom{2}} \vmom{2}^2}{4 m_1 m_2^3}
	\biggr) \nlqn
	+ \scpm{\vmom{1}}{\vspin{1}}\scpm{\vnun}{\vspin{2}}\biggl(
		-\frac{3 \scpm{\vnun}{\vmom{2}} \vmom{1}^2}{4 m_1^3 m_2}
		-\frac{15 \scpm{\vnun}{\vmom{1}}\scpm{\vnun}{\vmom{2}}^2}{4 m_1^2 m_2^2}
		+\frac{3 \scpm{\vnun}{\vmom{1}} \vmom{2}^2}{4 m_1^2 m_2^2} \nlqn
		-\frac{3 \scpm{\vnun}{\vmom{2}} \vmom{2}^2}{4 m_1 m_2^3}
	\biggr)
	+ \scpm{\vnun}{\vspin{1}}\scpm{\vnun}{\vspin{2}}\biggl(
		-\frac{3 \scpm{\vmom{1}}{\vmom{2}}^2{}}{8 m_1^2 m_2^2} 
		+\frac{105 \scpm{\vnun}{\vmom{1}}^2 \scpm{\vnun}{\vmom{2}}^2}{16 m_1^2 m_2^2} \nlqn
		-\frac{15 \scpm{\vnun}{\vmom{2}}^2 \vmom{1}^2}{8 m_1^2 m_2^2} 
		+\frac{3 \vmom{1}^2\scpm{\vmom{1}}{\vmom{2}}}{4 m_1^3 m_2}
		+\frac{3 \vmom{1}^2 \vmom{2}^2}{16 m_1^2 m_2^2}
		+\frac{15 \vmom{1}^2 \scpm{\vnun}{\vmom{1}}\scpm{\vnun}{\vmom{2}}}{4 m_1^3 m_2}
	\biggr) \nlqn
	+ \scpm{\vspin{1}}{\vspin{2}}\biggl(
		\frac{\scpm{\vmom{1}}{\vmom{2}}^2}{16 m_1^2 m_2^2}
		-\frac{9 \scpm{\vnun}{\vmom{1}}^2 \vmom{1}^2}{8 m_1^4}
		-\frac{5 \scpm{\vmom{1}}{\vmom{2}} \vmom{1}^2}{16 m_1^3 m_2} 
		-\frac{3 \scpm{\vnun}{\vmom{2}}^2\vmom{1}^2}{8 m_1^2 m_2^2} \nlqn
		-\frac{15 \scpm{\vnun}{\vmom{1}}^2 \scpm{\vnun}{\vmom{2}}^2}{16 m_1^2 m_2^2} 
		+\frac{3 \vmom{1}^2 \vmom{2}^2}{16 m_1^2 m_2^2}
		+\frac{3 \vmom{1}^2 \scpm{\vnun}{\vmom{1}}\scpm{\vnun}{\vmom{2}}}{4 m_1^3 m_2} \nlqn
		+\frac{9 \scpm{\vmom{1}}{\vmom{2}}\scpm{\vnun}{\vmom{1}}\scpm{\vnun}{\vmom{2}}}{16 m_1^2 m_2^2}
	\biggr)
\biggr] + (1\leftrightarrow2)\,,
\end{align}
Notice that from a combinatorial point of view there are 167 algebraically 
different possible contributions to the Hamiltonian for all objects
(written in terms of the canonical spin tensor), but 75 of them do not 
appear in the canonical representation used here.
The Hamiltonian is valid for any compact objects like black holes or neutron stars. 

The matter variables appearing in this Hamiltonian fulfill the standard Poisson bracket relations, namely
\begin{align}
 \{\hat{z}^i_a, \mom{a}{j}\} = \delta_{ij}\,, \;
 \{\spin{a}{i}{j}, \spin{a}{k}{\ell}\} = \delta_{ik} \spin{a}{j}{\ell} - \delta_{i\ell} \spin{a}{j}{k} 
	- \delta_{jk} \spin{a}{i}{\ell} + \delta_{j\ell} \spin{a}{i}{k}\,,
\end{align}
where the canonical spin tensor $\spin{a}{i}{j}$ is related to the canonical spin vector $\hat{\vct{S}}_{a}$ via
$\spin{a}{i}{j} = \varepsilon_{ijk} \hat{S}_{a\,(k)}$ and $\varepsilon_{ijk}$ is the Levi-Civita symbol. The appropriate Poisson brackets for the canonical spin vector are given by
\begin{align}
 \{\hat{S}_{a\,(i)}, \hat{S}_{a\,(j)}\} = \varepsilon_{ijk}\hat{S}_{a\,(k)}\,.
\end{align}
All other Poisson brackets are zero. Notice that the spin length
$\sqrt{\hat{S}_{a\,(i)} \hat{S}_{a\,(i)}}$ is a constant. (This is not necessarily
the case if the Hamiltonian would depend on angle-type variables describing the orientation
of the object.) Therefore each spin vector has only two dynamical degrees of freedom, which
are taken to form the spin part of the phase space, see, e.g., \cite{Damour:Jaranowski:Schafer:2008:1}.
The PN Hamiltonian $H$ can be used to get the time evolution of an arbitrary phase space 
function $A$ via
\begin{align}
 \frac{\text{d}A}{\text{d}t} &= \{A,H\} + \frac{\partial A}{\partial t}\,.
\end{align}

\section{Approximate Poincar\'e algebra}\label{sec:poincare}
As in \cite{Hartung:Steinhoff:2011:1} we utilize the (PN approximate) global Poincar\'e algebra
as a check of our calculation. Total linear momentum $\vct{P}$ and total angular momentum
$J^{ij} = - J^{ji}$ are still given by the same expressions, namely
\begin{align}
 \vct{P} = \sum_a \vmom{a}\,, \quad
 J^{ij} = \sum_a \left[\hat{z}^i_a \mom{a}{j} - \hat{z}^j_a \mom{a}{i} + \spin{a}{i}{j}\right]\,,
\end{align}
see also, e.g., \cite{Damour:Jaranowski:Schafer:2000, Damour:Jaranowski:Schafer:2008:1}.
For the contributions of the propagating field degrees of freedom see, e.g., \cite{Steinhoff:Wang:2009,Steinhoff:2011}.
As in \cite{Damour:Jaranowski:Schafer:2000, Damour:Jaranowski:Schafer:2008:1,Hartung:Steinhoff:2011:1} we
use an ansatz for the center-of-mass vector $\vct{G}$ at next-to-next-to-leading order spin(1)-spin(2) level,
since the integrals needed to be evaluated at this order are very hard to solve. 
This ansatz contains 86 unknown coefficients, whereas
the ansatz for the next-to-leading order case contains only 4 coefficients 
(if one fixes the $\vx{a}$-parts via the leading order Hamiltonian). At 
this order the integrals are also still solvable \cite{Steinhoff:Schafer:Hergt:2008}. 
In contrast, the next-to-next-to-leading order considered here will contain higher linear momentum powers than the next-to-leading order, 
which leads to a much higher number of irreducible algebraic quantities entering the center-of-mass vector. The $\vx{a}$-part 
of the center-of-mass vector can also be fixed by the next-to-leading order spin(1)-spin(2) Hamiltonian due to the $\{G^i, P^j\}$ 
Poisson bracket relation appearing in the Poincar\'e algebra. 
So there remain only the mentioned 86 coefficients, which were uniquely fixed by evaluating the $\{G^i, H\}$ Poisson 
brackets yielding 62 of them to be zero. The consistency of the solution obtained by evaluating the Poisson brackets
above was checked by evaluating the $\{G^i, G^j\}$ Poisson bracket relation and all other relations of the Poincar\'e algebra.

The center-of-mass vector at next-to-next-to-leading order spin(1)-spin(2) level is given by
\begin{align}
  \vct{G}^{\text{NNLO}}_{\text{SS}} & = 
	\frac{\gravthree^2}{\rel^3} ((\vnun \times \vspin{2})\times \vspin{1}) \biggl(\frac{17}{8} m_1 + m_2\biggr) \nln
	+ \frac{\gravthree}{\rel^2}\biggl[ \vmom{1} \biggl(
		-\frac{\scpm{\vnun}{\vspin{1}}\scpm{\vmom{2}}{\vspin{2}}}{4 m_1 m_2}
		+\frac{3 \scpm{\vnun}{\vspin{1}} \scpm{\vnun}{\vspin{2}} \scpm{\vnun}{\vmom{2}}}{4 m_1 m_2}
	\biggr) \nlqn
	+ 
	(\vnun \times \vspin{1}) \biggl(
		-\frac{((\vmom{1}\times\vmom{2})\,\vspin{2})}{4 m_1 m_2}
		+\frac{3((\vnun\times\vmom{1})\,\vspin{2})\scpm{\vnun}{\vmom{2}}}{4 m_1 m_2}
	\biggr) \nlqn
	-
	(\vmom{1} \times \vspin{1}) \frac{((\vnun \times \vmom{2})\,\vspin{2})}{8 m_1 m_2}
	-
	(\vmom{2} \times \vspin{1}) \frac{((\vnun \times \vmom{1})\,\vspin{2})}{4 m_1 m_2} \nlqn
	-
	((\vmom{1} \times \vspin{2})\times \vspin{1}) \frac{\scpm{\vnun}{\vmom{2}}}{4 m_1 m_2} 
	-
	((\vmom{2} \times \vspin{2})\times \vspin{1}) \frac{\scpm{\vnun}{\vmom{1}}}{4 m_1 m_2} \biggr]\nln
	+\frac{\vx{1}}{\rel} \biggl(
		\frac{2 \gravthree^2 (2 m_1 + m_2)}{\rel^3}\biggl[
			\scpm{\vspin{1}}{\vspin{2}}
			-2\scpm{\vnun}{\vspin{1}}\scpm{\vnun}{\vspin{2}}
		\biggr] \nlqn
		+\frac{\gravthree}{\rel^2}\biggl[
			-\frac{3 ((\vnun\times\vmom{1})\,\vspin{1})((\vnun\times\vmom{1})\,\vspin{2})}{2 m_1^2} \nlqn
			+\frac{3 ((\vnun\times\vmom{2})\,\vspin{1})((\vnun\times\vmom{1})\,\vspin{2})}{2 m_1 m_2}
			+\frac{3 ((\vnun\times\vmom{1})\,\vspin{1})((\vnun\times\vmom{2})\,\vspin{2})}{8 m_1 m_2} \nlqn
			-\frac{\scpm{\vmom{2}}{\vspin{1}}\scpm{\vmom{1}}{\vspin{2}}}{8 m_1 m_2}
			+\frac{\scpm{\vmom{1}}{\vspin{1}}\scpm{\vmom{2}}{\vspin{2}}}{4 m_1 m_2}
			+\frac{3 \scpm{\vmom{2}}{\vspin{1}}\scpm{\vnun}{\vspin{2}}\scpm{\vnun}{\vmom{1}}}{2 m_1 m_2} \nlqn
			-\frac{3 \scpm{\vnun}{\vspin{1}}\scpm{\vmom{1}}{\vspin{2}}\scpm{\vnun}{\vmom{1}}}{2 m_1^2}
			+\frac{3 \scpm{\vnun}{\vspin{1}}\scpm{\vmom{2}}{\vspin{2}}\scpm{\vnun}{\vmom{1}}}{4 m_1 m_2} \nlqn
			+\frac{3 \scpm{\vmom{1}}{\vspin{1}}\scpm{\vnun}{\vspin{2}}\scpm{\vnun}{\vmom{2}}}{4 m_1 m_2} \nlqn
			-\scpm{\vnun}{\vspin{1}}\scpm{\vnun}{\vspin{2}}\biggl\{
				\frac{15 \scpm{\vnun}{\vmom{1}} \scpm{\vnun}{\vmom{2}}}{4 m_1 m_2}
				+\frac{3 \scpm{\vmom{1}}{\vmom{2}}}{4 m_1 m_2}
			\biggr\} \nlqn
			+\scpm{\vspin{1}}{\vspin{2}}\biggl\{
				\frac{3 \scpm{\vnun}{\vmom{1}}^2}{2 m_1^2}
				-\frac{3 \scpm{\vnun}{\vmom{1}}\scpm{\vnun}{\vmom{2}}}{4 m_1 m_2}
				+\frac{\scpm{\vmom{1}}{\vmom{2}}}{8 m_1 m_2}
			\biggr\}
		\biggr]
	\biggr) + (1\leftrightarrow2)\,.
\end{align}
From this the boost vector $\vct{K} = \vct{G} - t \vct{P}$ can be obtained, which explicitly depends on time $t$.
%
%

\ifnotadp
\paragraph*{Acknowledgments}
\ack

\setlength{\bibsep}{0pt}
\providecommand{\href}[2]{#2}\begingroup\raggedright\endgroup

\fi

\bibliographystyle{adp}

\ifadp
\begin{acknowledgement}
\ack
\end{acknowledgement}
\input{refs_adp}
\fi

\end{document}